\documentstyle[preprint,aps,epsf]{revtex}
\tightenlines
\begin{document}
\setcounter{page}{0}

\draft

\title{Glueball Production in Peripheral Heavy-Ion Collisions}
\author{Alec J. Schramm \thanks{Email: {\it alec@phys.oxy.edu}}}
\address{Department of Physics, Occidental College, Los Angeles, CA 90041}

% Make date line blank
%\date{\today}

\maketitle
\thispagestyle{empty}

\begin{abstract} 
The method of equivalent quanta is applied both to photon-photon and, by analogy, to double pomeron exchange in heavy-ion collisions.  This Weizs\"acker-Williams approach is used to calculate production cross sections for the glueball candidate $f_J(1710)$ meson via  photon-photon and pomeron-pomeron fusion in peripheral heavy-ion collisions at both RHIC and LHC energies.  The impact-parameter dependence for total and elastic cross sections are presented, and are compared to results for proton-proton collisions. 
\end{abstract}

\section{Introduction}
Glueballs, bound color-singlet states of gluons, are a
well-known prediction of Quantum Chromodynamics (QCD) and have long been sought in the mesonic spectrum.
Three popular, glue-rich environments have been the focus of this search.
The most promising is radiative $J/\psi$ decay, $J/\psi\rightarrow\gamma G$, in which the $c\bar{c}$ pair decays to hadrons predominantly through two-gluon intermediate states.  A second possibility is the annihilation into gluons of one or two $q\bar{q}$ pairs in $p\bar{p}$ collisions (e.g., $p\bar{p}\rightarrow \pi^0 G$).  A third environment for glueball production is central $pp$ collisions, where gluon-gluon fusion processes should be abundant; glueball creation in this mode is expected to proceed via double pomeron exchange, with the pomeron presumed to be a multi-gluon color singlet.

A fourth, related venue for glueball production is in peripheral heavy ion collisions.  Although central collisions of heavy ions
at the Relativistic Heavy Ion Collider (RHIC) at Brookhaven and the Large Hadron Collider (LHC) at CERN  will form the frontlines in the search for a quark-gluon plasma, their large center-of-mass energies (as high as 100 GeV/nucleon at RHIC and 3.5 TeV/nucleon at the LHC) 
should allow access to the Regge region $s\gg |t|$ and thence to the production of glueballs in peripheral collisions by strong double diffractive scattering, i.e., pomeron-pomeron (${\cal PP}$) exchange.  

One popular glueball candidate is the $f_{J}(1710)$ meson, which is produced in radiative $J/\psi$ decay at a rate of about 1 part in $10^3$.  
There is some disagreement regarding the spin of this resonance\cite{pdg}, including evidence for two overlapping states obscuring one another\cite{bes}.
Nevertheless, lattice QCD calculations have provided evidence 
identifying the oft-suspected $f_{J}(1710)$ as a scalar glueball with little or no admixture of valence quark--anti-quark pairs \cite{ibm}.
Moreover, recent analysis of radiative $J/\psi$ decay indicates that
as a $q\bar{q}$ state, the rate of $f_{J}(1710)$ production is consistent only with $J=2$; for $J=0$ the production rate is too high for a pure $q\bar{q}$ state, but is compatible with a glueball or mixed $q\bar{q}$-glueball\cite{cf,cfl}.
This strengthens the claim for a ground state scalar glueball in the 1700 MeV range.  
As such, the production and study of the $f_J(1710)$ could provide an important testing ground for QCD.  

In this paper we discuss the production of $f_{J=0}(1710)$ mesons via pomeron-pomeron and photon-photon interactions in peripheral heavy ion collisions at both RHIC and the LHC.  
In addition to pomerons,
these facilities will also endow the colliding nuclei with large virtual photon luminosities; since gluons are charge neutral, however, $\gamma\gamma$ production of glueballs will not occur to leading order.  Nonetheless, such large photon luminosities raise the prospect of studying electromagnetic particle production via two photon interactions \cite{baur}, and despite the expectation that the contribution from the $\gamma\gamma$ mode is small, it will interfere constructively with the ${\cal PP}$ process. Perhaps more importantly, just as one can calculate such $\gamma\gamma$ processes in an equivalent photon approximation, one can construct by analogy an ``equivalent pomeron approximation'' for ${\cal PP}$ exchange\cite{us2}.
Thus we begin in Sec. II with a review of the method of virtual quanta as applied to photons and pomerons.  Although the equivalent photon approximation is well known, we review the main features in order to facilitate comparison with equivalent pomerons; the presentation will follow the outline of earlier work on the subject \cite{us2,us}.
We also discuss the impact parameter dependence of the two-photon and two-pomeron processes, complementing earlier work by Natale \cite{natale}.  In Sec. III we present our results for both photon-photon and pomeron-pomeron interactions, including the effects of nuclear absorption.

\section{Equivalent Photons and Pomerons}
We consider the production of a state $X^0$ of large invariant mass $M_X$ by the exchange of two virtual quanta from the scattering nuclei.  Let $k_i^2, \omega_i~(i=1,2)$ be the invariant masses and energies of the exchanged quanta, $p_i, p^\prime_i$ the four-momenta of the two nuclei before and after the collision, and $M_i$ their invariant masses.  We assume that the nuclei remain in their ground state, i.e., the collision is elastic with respect to the scattering nuclei.  The relevant four-momentum can then be expressed as
\begin{eqnarray}
p_i &=& (E_i, p_{i\parallel},\bf{0}_\perp) \\
p_i^\prime &=& (E_i^\prime, p_{i\parallel}^\prime,-\bf{k}_{i\perp}) \\
k_i &=& (\omega_i, k_{i\parallel},\bf{k}_{i\perp}) ,
\end{eqnarray}
where the notation for the various particle momenta is explained in Figure 1.  In the high-energy limit, $E_i\gg\omega_i$, we have
\begin{eqnarray}
\omega_i &=& E_i - E_i^\prime \equiv x_iE_i \\
k_{i\parallel} &=& p_{i\parallel} - p_{i\parallel}^\prime \approx \frac{dp_{i\parallel}}{dE_i}(E_i - E_i^\prime) = \frac{E_i}{p_{i\parallel}} x_iE_i ~,
\end{eqnarray}
which together give
\begin{eqnarray}
\omega_i^2 - k_{i\parallel}^2 \approx \omega_i^2\left(1 - \frac{E_i^2}{p_{i\parallel}^2}\right) \approx - x_i^2 M_i^2 ~,
\end{eqnarray}
leading to the useful relation
\begin{eqnarray}
-k_i^2 = x_i^2 M_i^2 + \bf{k}_{i\perp}^2 ~.
\end{eqnarray}

In the spirit of the equivalent photon approximation \cite{them}, for ion beams $A$ and $B$ with squared center of mass energy $s$ we write the cross section for either photon-photon or pomeron-pomeron fusion to a neutral state $X$ in the form 
\begin{eqnarray}
\sigma^{^{VV}}_{AB} = 
\int~dx_1dx_2~f_{_V}^A(x_1)f_{_V}^B(x_2)
\hat{\sigma}_{_{VV}}^X(x_1x_2s) ~.
\end{eqnarray}
In this expression $f_{_V}(x)$ is the distribution function for finding a quantum $V$ with energy fraction $x$ in the nucleus.
The production cross section for the subprocess $VV\rightarrow X$ with squared center-of-mass energy $x_1x_2s$ is given by\cite{brodsky}
\begin{eqnarray}
\hat{\sigma}_{_{VV}}^{X} = (2J+1)~\frac{8\pi^2}{M_{_X}}~
      \Gamma_{_{X\rightarrow VV}}~\delta(x_1x_2s - M_{_X}^2)~,
\end{eqnarray}
where $J$ is the spin of the state $X$ and $\Gamma_{_{X\rightarrow VV}}$ is the partial two-$V$ decay width of $X$.

For the photon case ($V=\gamma$), the $\gamma$-nucleus vertex is given by $ZeF_A(t)$, where  $F_A$ is the elastic nuclear form factor and $t$ is the invariant four-momentum exchanged.  Then the distribution function $f_\gamma$ for a fast moving nucleus of charge $Z$, mass $M$, and nucleon number $A$ is \cite{us}
\begin{eqnarray}
f_\gamma(x) &=& \frac{(Ze)^2}{\pi x} \int_0^\infty \frac{d^2k_\bot}{(2\pi)^2} ~k_\bot^2~ \frac{\,|F_A(-k^2)|^2}{(-k^2)^2} \\
              &=& \frac{Z^2\alpha}{\pi x}\int_0^\infty d(k_\bot^2) ~k_\bot^2~\frac{\,|F_A(x^2M^2 + k_\bot^2)|^2}{(x^2M^2 + k_\bot^2)^2} ~,
\end{eqnarray}
where we have used Equation (7).
Approximating $F(\vec{k}^2)$ by the gaussian 
\begin{eqnarray}
F(\vec{k}^2) = e^{-\vec{k}^2/2Q_0^2}~,
\end{eqnarray}
we see that the form factor imposes a cutoff $xM/Q_0 \sim xMR \,\,\raisebox{-0.13cm}{$\stackrel{\textstyle<}{\textstyle\sim}$}\,\, 1$, where $R \sim 1/Q_0$ is the nuclear radius.  Thus for identical nuclei, a heavy state of invariant mass $M_X$ can be produced as long as $M_X^2 = x_1x_2s \,\,\raisebox{-0.13cm}{$\stackrel{\textstyle<}{\textstyle\sim}$}\,\, s/M^2R^2 = (2\gamma/R)^2$.  This is essentially the condition for coherence: the photon wavelength must be larger than the Lorentz--contracted nuclear radius.
Coherence leads to the factor of $Z^2$ in $f_\gamma$ (compounded to $Z^4$ in the cross section), rendering electromagnetic interactions of high-$Z$ ions an effective tool for the production of heavy neutral particles.

Turning next to the pomeron ($V={\cal P}$), we distinguish the exclusive process of Figure 1 from the inclusive process shown in Figure 2.  Considering the pomeron as a two gluon ladder, the exclusive case requires one of the gluons of each pomeron to fuse.  Such a process warrants a full gluon ladder calculation, and almost certainly yields a small rate.  By contrast, inclusive production puts no restrictions upon the two remaining gluons, which are soft on average, and a ladder calculation may not be necessary.  Moreover, the inclusive rate should be larger than that for exclusive production, and is the process that will be seen first.  We thus proceed for pomerons by considering only inclusive production.

As such, we are interested in the so-called ``soft pomeron'', considered to be an effective description of a highly correlated gluon exchange with a structure determined by nonperturbative properties of the QCD vacuum.  One begins with the Regge trajectory
\begin{eqnarray}
\alpha_{_{\cal P}}(t) = 1 + \epsilon + \alpha_{_{\cal P}}^\prime t,
\end{eqnarray}
where $\epsilon = 0.085$ and $\alpha_{_{\cal P}}^\prime = 0.25$ GeV$^{-2}$. 
This trajectory has an intercept close to 1, suggesting that the pomeron behaves like a spin-one boson in processes in which the exchanged four-momentum squared $k^2=t$ is small.  Pomeron exchange can thus be described as an isoscalar ``photon'', with propagator at high c.m. energy given by \cite{DL1},
\begin{eqnarray}
|D_{_{\cal P}}(t=-\vec{k}^2;s)| = (s/m^2)^\epsilon~e^{-r_0^2\vec{k}^2}
\end{eqnarray}
where $m$ is the Regge scale parameter and $r_0$ is the pomeron range parameter (see Appendix),
\begin{eqnarray}
r_0^2 = \alpha_{_{\cal P}}^\prime\ln{(s/m^2)}~.
\end{eqnarray}
Following Donnachie and Landshoff \cite{DL2}, we denote the nucleon-pomeron coupling by
\begin{eqnarray}
\beta_{NP} = 3\beta_0F_N(-t) ~,
\end{eqnarray}
where the quark-pomeron coupling $\beta_0$ is
\begin{eqnarray}
\beta_0 = 1.8~\mbox{GeV}^{-1} 
\end{eqnarray}
and $F_N(-t)$ is the isoscalar magnetic nucleon form factor.  
For nucleus-pomeron interaction, we appeal to the additivity of the total nucleon-nucleon cross sections and replace $\beta_0$ by $A\beta_0$
and $F_N(-t)$ by the elastic nuclear form factor $F_A(\vec{k}^2)$.  Of course, a linear dependence of the nucleus-pomeron coupling on $A$ is too strong:  While it may be appropriate for a hard process involving at least 100 GeV, for the softer process considered here shadowing will moderate the coupling to scale like $A^\delta$ with $\delta<1$. For small $x$, where the method of equivalent quanta applies, $\delta\sim 0.7-0.8$ \cite{busza}.
We will nevertheless use the linear coupling for simplicity and apply this suppression to the resulting cross sections.

For inclusive glueball production, the distribution function for finding a pomeron in the nucleus with energy fraction $x$ is then given by \cite{us2,DL2}
\begin{eqnarray}
f_{_{\cal P}}(x) = \left(\frac{3A\beta_0Q_0}{2\pi}\right)^2
                     \left(\frac{s^\prime}{m^2}\right)^{2\epsilon}
                     \frac{1}{x}~e^{-x^2M^2/Q_0^2} ~,
\end{eqnarray}
where $s^\prime$ denotes the invariant subprocess with which the pomeron participates, i.e., $s^\prime \equiv s_1 \approx x_2s$ for the pomeron emitted by nucleus 1, $s^\prime \equiv s_2 \approx x_1s$ for the pomeron emitted by nucleus 2.   

Being concerned with inclusive production, an additional step is required between the subprocess cross section $\hat{\sigma}$ called for in Equation (8) and the exclusive form advocated in Equation (9).  Specifically, we replace Equation (9) with
\begin{eqnarray}
\hat{\sigma}_{_{\cal  PP}}^{\rm incl} = 
  \int~dz_1dz_2~G_{_{\cal P}}(z_1)G_{_{\cal P}}(z_2)
\tilde{\sigma}_{_{gg}}^X(z_1z_2 s_{_{\cal PP}}) ~,
\end{eqnarray}
where $G_{_{\cal P}}(z)$ is the gluon structure function of the pomeron and $s_{_{\cal PP}} \equiv x_1x_2s$ is the squared center-of-mass energy of the two-${\cal P}$ system.  
The gluonic glueball production cross section needed in Equation (19) is now given by
\begin{eqnarray}
\tilde{\sigma}_{_{gg}}^{X}(s) = (2J+1)~\frac{\pi^2}{8M_{_X}}~
      \Gamma_{_{X\rightarrow gg}}~\delta(s - M_{_X}^2)~,
\end{eqnarray}
which differs from Equation (9) only by a factor of $\left(\frac{1}{8}\right)^2$ due to averaging over initial color states. 
As for $G_{_{\cal P}}(z)$,
experimental investigation into the partonic structure of the pomeron indicates that most of the pomeron momentum is carried by hard gluons \cite{zeus}, favoring a structure function of the form
\begin{eqnarray}
G_{_{\cal P}}(z) = 6(1-z) ~.
\end{eqnarray}
This, however, raises other concerns.  As the mass $M_X$ of the produced particle decreases, a structure function approach to the glue content of the pomeron becomes invalid: without a large momentum scale one cannot fully justify a parton model formalism.  (Indeed, structure functions such as  (21) are obtained from data with large $p_T$ jets which serve as the scale for a perturbative approach.)   Just how small $M_X$ must be before the approach is no longer reliable is an open question, but it seems not unreasonable to be suspicious for  $M_X\simeq1.7$ GeV.   Thus the calculation presented here is not intended to be truly predictive; rather,  it is offered as an order of magnitude estimate in a first attempt to see if the experiment makes sense. 
In order to get an idea how sensitive the results are to the choice of structure function, the calculation was performed not only using $G_{_{\cal P}}(z)$ as in (21), but with two invented structure functions as well, one softer  the other harder than (21),
\begin{eqnarray}
G^\prime_{_{\cal P}}(z) &=& 12(1-z)^2 \\
G^{\prime\prime}_{_{\cal P}}(z) &=& \frac{15}{4}(1-z)^{1/2}~.
\end{eqnarray}

Although we have now assembled the expressions required by Equation (8), a practical consideration remains.  In order to obtain a clean glueball signal, it is important that the cross section remain viable out to impact parameters greater than roughly twice the nuclear radii---in other words, that the heavy ions engage in peripheral collisions. To calculate the impact parameter dependence of the cross section, we write the total $VV$ exchange cross section (8) in the impact parameter representation and fold it with the probability that no inelastic interaction takes place other than double-$V$ exchange.  The dependence of the cross section on the impact parameter $b$ is then found by integrating the squared matrix element over all spacetime coordinates save for the transverse distance $b$ between nuclei \cite{us,us2}.  

The key ingredient in this recipe is the effective $V$-$X$ coupling, which depends on the spin and parity of $X$. If we denote the vertex by $\Gamma_{\mu\nu}$ and in the high-energy limit keep only the transverse polarizations of the $V$'s, the coupling of a scalar $X$ to the coherent nuclear current can be expressed in the form  
\begin{eqnarray}
(p_1 + p_1^\prime)^\mu\Gamma_{\mu\nu}(p_2 + p_2^\prime)^\nu
    \approx 4\sqrt{2}~\frac{{\bf k}_1\cdot{\bf k}_2}{x_1x_2}~ 
     \overline{M}_{_{VV}}(k_1,k_2)~,
\end{eqnarray}
where $\overline{M}_{_{VV}}$ is the spin-averaged invariant matrix element for the process $VV\rightarrow X$.  The factor ${\bf k}_1\cdot{\bf k}_2$ would be replaced $|{\bf k}_{1\bot}\times {\bf k}_{2\bot}|$ for a pseudoscalar $X$.  The different forms of $\Gamma_{\mu\nu}$ for a number of other possible spin-parity states $X$ can be easily deduced from the list of two-photon helicity amplitudes given in \cite{poppe}. 

This procedure, together with the Gaussian form factor (12), results in  straightforward calculations for the differential cross sections.  For completeness, we display below expressions for both $J=0$ and $2$. (Note that for $J=2$, the $h=\pm 2$ helicity states in $\gamma\gamma\rightarrow X$ have been observed to dominate \cite{exp}, a phenomenon that can be understood in the context of tensor meson dominance \cite{suzuki}.  
Thus when treating the $f_J(1710)$ as a $2^+$ state, the helicity zero contributions can be neglected, greatly simplifying $\Gamma_{\mu\nu}$.)

For $\gamma\gamma$ exchange, we get the general expression
\begin{eqnarray}
\frac{d\sigma^{\gamma\gamma\rightarrow f_J}_{AB}}{d^2b} = 
  \frac{1}{\pi}\left(\frac{Z_1Z_2\alpha}{\pi}\right)^2 
 \int\frac{dx_1}{x_1}\frac{dx_2}{x_2}~\hat{\sigma}_{\gamma\gamma}^{f_J}(x_1x_2 s)
 \int~dq~q~J_0(qb)~e^{-q^2/2Q_0^2} ~\phi_J(x_1,x_2;q)
\end{eqnarray}
where we have defined
\begin{eqnarray}
\phi_{J=0} &=&    \left(A_1(x_1) - \frac{q^2}{4Q_0^2}B_0(x_1)\right) 
   \left(A_1(x_2) - \frac{q^2}{4Q_0^2}B_0(x_2)\right) + \nonumber \\
    & &  \hspace*{2in} \left(B_1(x_1) - \frac{}{}A_1(x_1)\right)
                         \left(B_1(x_2) - \frac{}{}A_1(x_2)\right)  \\
\phi_{J=2} &=& \left(B_1(x_1) - \frac{q^2}{4Q_0^2}B_0(x_1)\right) 
                \left(B_1(x_2) - \frac{q^2}{4Q_0^2}B_0(x_2)\right) ~,
\end{eqnarray}
with
\begin{eqnarray}
A_n(x) &=& e^{-x^2M^2/Q_0^2}\int_0^\infty~d\xi~\xi^n~\frac{e^{-\xi}}{b}
   \left[\sqrt{\frac{a}{a-b}} - 1 \right] \\
B_n(x) &=& e^{-x^2M^2/Q_0^2}\int_0^\infty~d\xi~\xi^n~  
   \frac{e^{-\xi}}{\sqrt{a(a-b)}}  ~,
\end{eqnarray}
and
\begin{eqnarray}
a\equiv \left(\xi + \frac{q^2}{4Q_0^2} + \frac{x^2M^2}{Q_0^2}\right)^2~,
            ~~~~ b\equiv\xi q^2/Q_0^2 ~.
\end{eqnarray}

The expressions are simpler for ${\cal PP}$ exchange, allowing much of the integration to be performed analytically; for identical nuclei, we find
\begin{eqnarray}
\sigma^{\cal PP}_{AA} = \left(\frac{3A\beta_0Q_0}{2\pi}\right)^4 
           \frac{\pi^2 \Gamma_{_{X\rightarrow gg}}}{8M_{_X}^3}
           \int~d\xi \left(\frac{s M_{_X}^2}{m^4\xi}\right)^{2\epsilon}
           \int~\frac{dz}{z}~G_{_{\cal P}}(z) G_{_{\cal P}}(\xi/z)
           K_0\left(\frac{2M_{_X}^2M^2}{sQ_0^2\xi}\right) ~,
\end{eqnarray}
where $\xi = M_{_X}^2/x_1x_2s$. 
Since the pomeron propagator is a gaussian, the choice of a gaussian nuclear form factor allows the impact parameter dependence to be determined exactly.  
The differential cross section for $J=0$ is found to be
\begin{eqnarray}
\frac{d\sigma^{{\cal PP}\rightarrow f_{J=0}}_{AA}}{d^2b} = 
   \frac{Q^{\prime 6}_0 b^4}{16\pi}~e^{-b^2 Q^{\prime 2}_0/2}~ \sigma^{\cal PP}_{AA}~,
\end{eqnarray}
whereas 
\begin{eqnarray}
\frac{d\sigma^{{\cal PP}\rightarrow f_{J=2}}_{AA}}{d^2b} = 
   \frac{Q^{\prime 2}_0}{4\pi}\left(\frac{1}{8}b^4 Q^{\prime 4}_0 +1\right)~
      e^{-b^2 Q^{\prime 2}_0/2}~ \sigma^{\cal PP}_{AA}~
\end{eqnarray}
for $J=2$.
In both these expressions, we have defined
\begin{eqnarray}
Q^{\prime -2}_0 \equiv Q_0^{-2} + 2r_0^2 ~.
\end{eqnarray}

\section{Glueball Production}
In order to derive numerical values for these cross sections, estimates for the two-photon and two-gluon decay widths are needed.  Determining these widths, however, depends upon whether the state is pure quarkonium, pure glue, or even a $q\bar{q}$--glue hybrid.  
For a pure quarkonium state, the two-photon and two-gluon widths can be related to leading order by replacing the photon coupling
$\alpha$ with the gluon coupling $\alpha_s$,
\begin{eqnarray}
\Gamma_{_{(q\bar{q})\rightarrow\gamma\gamma}} = \frac{9e_q^4}{2}\left(\frac{\alpha}{\alpha_s}\right)^2 
\Gamma_{_{(q\bar{q})\rightarrow gg}} ~,
\end{eqnarray}
where 9/2 is the color factor and $e_q$ is the relevant quark charge.  
One can estimate $\Gamma_{_{(q\bar{q})\rightarrow gg}}$ from the measured total $f_J(1710)$ width of 140 MeV \cite{pdg} and the expectation that the $(q\bar{q})\rightarrow gg$ branching ratio is of order $\alpha_s^2$ \cite{cf}.
The numerical value of $\Gamma_{_{(q\bar{q})\rightarrow\gamma\gamma}}$ is then estimated to be $\sim$ 1 keV.

Glue states are more difficult. Whereas one expects a $(q\bar{q})\rightarrow gg$ branching ratio of order $\alpha_s^2 \simeq 0.2$, a branching ratio of order 1 is anticipated for $G\rightarrow gg$.
In fact, even a two-gluon branching ratio as small as $\sim 1/2$ would make it difficult to classify the state as pure quarkonium\cite{cf,cfl}. 
We rely on the analysis of Close, Farrar, and Li \cite{cfl} and
take the branching ratio for $f_0(1710)\rightarrow gg$ to be $\approx 0.52$ (a value which is in accord with the analysis of reference \cite{cf}), which yields $\Gamma{(f_0\rightarrow gg)}\approx 70$ MeV. 
For the two-$\gamma$ decay width, we use the results of Kada, Kessler, and Parisi \cite{kada} who calculated $\Gamma{(G\rightarrow\gamma\gamma)}$ using a nonrelativistic gluon bound-state model.  Adapted for the $f_0(1710)$ and with a $f_0(1710)\rightarrow K\bar{K}$ branching ratio of $\sim 1/2$, their expressions yield $\Gamma{(f_0\rightarrow\gamma\gamma)} \approx 4$ eV. 
Our estimates for the various widths are summarized in Table 1.

We have used the two-gluon widths to calculate the production of a $f_{J=0}(1710)$ glueball via double-photon and double-pomeron exchange in heavy-ion collisions at LHC and RHIC energies. 
We consider $^{208}$Pb nuclei, and use the Gaussian form factor (12) with $Q_0\approx 60$ MeV\cite{dez}. 
Our calculations then yield a total cross section of $\sigma_{AA}^{\rm tot}(\gamma\gamma\rightarrow f_0(1710)) = 2.5~\mu$b at the LHC, $\sigma_{AA}^{\rm tot}(\gamma\gamma\rightarrow f_0(1710)) = 85$ nb at RHIC.  By contrast, the corresponding cross sections for the ${\cal PP}$ production mode are orders of magnitude higher: some 2160 b and 73 b at the LHC and RHIC, respectively. 

It is useful to compare these results with those originating from proton-proton rather than heavy ion interactions.  Noting that $Q_0$ scales like the inverse nuclear radius, i.e., as $A^{-1/3}$, we see from (31) that $\sigma^{\cal PP}_{AA} \sim A^{4/3}$.
Although the loss of large $A$ in proton-proton collisions will certainly lower the pomeron-pomeron cross section, the additional luminosity and energy available to proton beams at RHIC ($\sim 10^{32}~{\rm cm}^{-2}~{\rm s}^{-1}$, 250 GeV) should offset this expected decrease.  For simplicity of comparison, we have approximated the usual nucleon dipole form factor with the Gaussian of (12), taking $Q_0^2 = 0.71/4 = 0.1775~{\rm GeV}^2$.  The resulting cross section for the ${\cal PP}$ production of $f_0(1710)$ is 151 $\mu$b which, at the increased RHIC luminosity, is competitive with production in heavy ion collisions.  Not surprisingly, the extra energy and luminosity which result in large event rates for ${\cal PP}$ fusion is of little consequence for $\gamma\gamma$ fusion, since $\sigma^{\gamma\gamma}_{AA} \sim Z^{4}$.  Indeed, calculation shows the resulting cross section to be at the $0.5$ pb level.

These expectations, of course, are overly optimistic 
not only because they assume 100\% detection efficiency, but because
they do not account for the effects of inelastic nuclear scattering.  The majority of the inelastic events is expected to occur at small values of the impact parameter $b$;  indeed, the elastic nature of the interaction is maintained only in those collisions in which the two nuclei pass by each other.  Thus it is important to verify that a significant portion of the $\gamma\gamma$ and ${\cal PP}$ cross sections extends out to relatively large impact parameters.  For comparison, we have included inelastic scattering effects in two different ways.  One is by applying a geometric cutoff at a minimum impact parameter of $2R$, where $R$ is the nuclear radius ($R\approx 7.1$ fm for $^{208}$Pb).  A more realistic approach accounts for inelastic scattering effects using the Glauber approximation, with an absorption factor \cite{glauber}
\begin{eqnarray}
\frac{d\sigma^{\rm el}_{{\scriptscriptstyle A}{\scriptscriptstyle B}}}{d^2b} = 
   \frac{d\sigma_{{\scriptscriptstyle A}{\scriptscriptstyle B}}}{d^2b}~\exp{[-AB~T_{{\scriptscriptstyle A}{\scriptscriptstyle B}}(b)~\sigma_0]} ~,
\end{eqnarray}
where $\sigma_0$ is the total nucleon-nucleon cross section.  (We have used the formula of Amaldi \cite{amaldi} to extrapolate $\sigma_0$ to RHIC and LHC energies.)
For simplicity, we use the same gaussian form factor (12) to calculate the profile function \cite{yang}
\begin{eqnarray}
T_{{\scriptscriptstyle A}{\scriptscriptstyle A}}(b) = \int\frac{d^2Q}{(2\pi)^2}~F_{\scriptscriptstyle A}(Q^2)~F_{\scriptscriptstyle A}(Q^2)~e^{i\vec{Q}\cdot\vec{b}}  = \frac{Q_0^2}{4\pi}~e^{-Q_0^2b^2/4} 
~.
\end{eqnarray}
For $^{208}$Pb, this approach yields ${\cal PP}$ elastic cross sections of approximately 95 mb at the LHC and 8.2 mb at RHIC. Including shadowing effects further suppresses these cross sections: letting $A\rightarrow A^{\delta}$ with $\delta\sim 0.8$ lowers the ${\cal PP}$ cross sections by about an order of magnitude.  For proton scattering at RHIC, the ${\cal PP}$ cross section falls to 121 $\mu$b, and thus may be the more fruitful experimental approach. 
Our numerical results are summarized in Tables 2 and 3.  The dependence of the differential cross sections on impact parameter at RHIC is shown in Figure 3, where the dotted curves show the suppressing effects of absorption in the Glauber approximation;  
Figure 4 shows the impact parameter dependence for proton-proton collisions at RHIC.   

The pomeron results presented in Table 3 all employ the structure function $G_{_{\cal P}}$ of Equation (21). As discussed in the previous section, however, without a large momentum scale one cannot fully justify the use of the parton model formalism.  In order to gauge the sensitivity of the results on the choice of structure function, both $G^{\prime}_{_{\cal P}}$ and $G^{\prime\prime}_{_{\cal P}}$ of Equations (22) and (23) were also considered.  The results are shown in Figure 5, where the differential cross section which emerges using  $G_{_{\cal P}}$ is given by the solid curve, and is to be compared with the cross sections found using  $G^{\prime}_{_{\cal P}}$ and $G^{\prime\prime}_{_{\cal P}}$, given by the dashed and dotted curves, respectively.  The predicted cross sections are all found to be well within an order of magnitude of one another.  

As can be deduced from Equation (34), the difference between $Q^\prime_0$ and $Q_0$ corresponds to the folding of the nuclear density distribution with the range of the pomeron field.  This behavior is evident in Figure 3: the double-pomeron exchange process essentially depends on the magnitude of the geometrical overlap of the two nuclei during collision.
This is more readily appreciated in Figure 6, which shows the ratio of ${\cal PP}$ to $\gamma\gamma$ cross sections; an approximate exponential fall-off is evident, reflecting the pomeron's short range. 
Moreover, the pomeron contribution is more strongly suppressed because the probability for an elastic collision becomes large only for $b \,\,\raisebox{-0.13cm}{$\stackrel{\textstyle>}{\textstyle\sim}$}\,\, 16$ fm. 
 
Of course, it is not yet settled whether the $f_J(1710)$ meson is actually a glueball, and one must consider the experimental feasibility of using peripheral heavy-ion collisions to determine its nature.
Although the analysis of Close, Farrar, and Li \cite{cfl} rules out a glueball interpretation for $J=2$, our results are easily extended to a $J=2$ quarkonium state since the total cross section simply scales by the ratio of the two-body decay widths and by the $2J+1$ factor of Equation (20).  (Note that although the impact parameter dependence can in principle distinguish between the $J=0$ and $J=2$ cases, the difference is negligible once absorption is accounted for.)
For $J=0$, one must be concerned not only about a putative $\gamma\gamma\rightarrow G$ process being overwhelmed by conventional $q\bar{q}$ states, but also the inundation of both of these by the ${\cal PP}$ process.
Note however that as a probe only of the electrically charged $q\bar{q}$ component, the $\gamma\gamma$ cross section can be varied by changing the nuclear charge $Z$ at fixed energy.
With $gg$ decay widths which differ by a factor of 2.5 (and perhaps by as much as a factor of 5 should the $f_{J}(1710)$ be pure glue), one should therefore be able to distinguish a quarkonium from a glueball signal.  
As for concerns of pomeron fragmentation, the hard glue component of the pomeron carries $80-90\%$ of the pomeron's momentum, leaving only $10-20\%$ to fragment.  Moreover, the need for small $x$ requires that the two pomerons which form the state be soft: to make a 1700 MeV state each pomeron must have less than $\sim 1000$ MeV longitudinal momentum, of which 850 MeV goes into the produced state and 150 MeV fragments.  Thus any fragmentation must yield a 300 MeV, isospin zero state---most likely pion pairs.  There seems little reason to expect that this will swamp a glueball signal.

In conclusion, we find that peripheral collisions of heavy-ions---and, at RHIC, of protons---may be a fruitful approach to the production and study of glueballs.  For the $f_{J}(1710)$, the extent to which the measured production cross section agrees with results presented here may serve as a measure of the status of its candidacy; application to other glueball candidates is straightforward.

\section*{Appendix}
An essential difference between the production of states of high invariant mass by elastic two-photon exchange and that by double-pomeron exchange is that the electromagnetic interaction is long-ranged, whereas the strong interactions have a range of typically 1 fm.  In order to understand how these features are contained in the scattering formalism employed here, we briefly examine the range of the various interactions \cite{us2}.  In momentum space the distribution of virtual quanta around their source is given by
\begin{eqnarray}
\tilde{n}(t) = \Gamma(t)|D(t)| ~,
\end{eqnarray}
where $t$ is the invariant four-momentum exchanged, $\Gamma$ is the vertex function of the source, and $D$ is the propagator of the virtual quanta.  For virtual photons emitted by a nucleus, we have
\begin{eqnarray}
D_\gamma(t) = -1/t ~, ~~~~~~~~~\Gamma_\gamma(t) = ZeF_A(-t)~,
\end{eqnarray}
where $F_A$ is the elastic nuclear form factor.  For the pomeron, we have \cite{DL1,DL2}
\begin{eqnarray}
D_{_{\cal P}}(t) = \frac{(s/m^2)^{\alpha_{_{\cal P}}\!(t)-1}}{\sin\frac{1}{2}\pi\alpha_{_{\cal P}}\!(t)}~e^{-\frac{i}{2}\pi\alpha_{_{\cal P}}\!(t)}~, 
~~~~~~~~~\Gamma_{_{\cal P}}(t) = 3A\beta_0F_A(-t)~,
\end{eqnarray}
where $\alpha_{_{\cal P}}\!(t)$ is the Regge trajectory of the pomeron, and $3\beta_0$ is the pomeron-nucleon coupling constant.  The $\sin({\scriptstyle\frac{1}{2}}\pi\alpha_{_{\cal P}}\!)e^{\frac{i}{2}\pi\alpha_{_{\cal P}}}$ term is part of the so-called ``signature factor'' which expresses the different properties of the pomeron under $C$ and $P$ conjugation.  
We are interested here in the spatial distribution of the virtual quanta in the nuclear rest frame, where we have $t=-\vec{k}^2$.  Note that the pomeron propagator depends on the total scattering energy $s$, i.e., the pomeron distribution seen by the other nucleus (or any other probe) is a function of energy even in the rest frame of the source.  

The spatial distribution $n(r)$ is obtained from (36) by a Fourier transformation.  For the virtual photon field we find the well-known result,
\begin{eqnarray}
n_\gamma(\vec{r}) &=& \int\frac{d^3k}{(2\pi)^3}~e^{i\vec{k}\cdot\vec{r}}~
                                 \tilde{n}_\gamma(-\vec{k}^2) \nonumber \\
                  &=& \frac{Ze}{4\pi}\int d^3r^\prime~\frac{\rho(\vec{r}^{\,\prime})}{|\vec{r}-\vec{r}^{\,\prime}|}~,
\end{eqnarray}
where $\rho(\vec{r})$ is the (nuclear) charge density distribution.  For the pomeron distribution we find
\begin{eqnarray}
n_{_{\cal P}}\!(\vec{r})
        &=& \int\frac{d^3k}{(2\pi)^3}~e^{i\vec{k}\cdot\vec{r}}~
                     \tilde{n}_{_{\cal P}}\!(-\vec{k}^2) \nonumber \\
        &=& 3A\beta_0\int d^3r^\prime~\rho(\vec{r}^{\,\prime})\int \frac{d^3k}{(2\pi)^3}~e^{i\vec{k}\cdot(\vec{r}-\vec{r}^{\,\prime})}
             \frac{(s/m^2)^{\epsilon-\alpha_{_{\cal P}}^\prime\!\vec{k}^2}}{\sin\frac{1}{2}\pi(1+\epsilon-\alpha_{_{\cal P}}^\prime\!\vec{k}^2)}~,
\end{eqnarray}
where we have used Eq. (13).  At very high c.m. energy the numerator of the pomeron propagator falls off very rapidly with $\vec{k}^2$, with an exponential slope given by $\alpha_{_{\cal P}}^\prime\!\ln(s/m^2)$.  We can therefore neglect the $\vec{k}^2$ dependence of the signature factor and approximate
\begin{eqnarray}
\sin[{\scriptstyle\frac{1}{2}}\pi(1+\epsilon-\alpha_{_{\cal P}}^\prime\!\vec{k}^2)] \approx \cos {\scriptstyle\frac{1}{2}}\pi\epsilon \approx 1 ~.
\end{eqnarray}
Defining the pomeron range parameter $r_0$ as 
\begin{eqnarray}
r_0^2 = \alpha_{_{\cal P}}^\prime\!\ln(s/m^2) ~,
\end{eqnarray}
we obtain:
\begin{eqnarray}
n_{_{\cal P}}\!(\vec{r}) \approx \frac{3A\beta_0}{(2r_0\sqrt{\pi})^3}~(s/m^2)^\epsilon~
                              \int d^3r^\prime~\rho(\vec{r}^{\,\prime})~exp\left(-\frac{(\vec{r}-\vec{r}^{\,\prime})^2}{4r_0^2}\right) ~.
\end{eqnarray}
We thus find that the pomeron distribution corresponds to the nuclear density distribution folded with a gaussian of range $r_0$.  At LHC energy for $^{208}$Pb nuclei, with $m\approx 1$ GeV and $\alpha_{_{\cal P}}^\prime = 0.25$ GeV$^{-2}$, we find $r_0(100$ GeV/u$) = 0.45$ fm.  We conclude that the distribution of virtual pomerons falls off very rapidly at the nuclear surface, with a mean range less than that of the nuclear density itself.  

\acknowledgements
\noindent
I would like to thank Professor B. M\"uller for useful discussions. 
This work was supported by the Research Corporation.

\pagebreak

\section*{Tables}
%\large

\newcounter{num}
\begin{list}%
{Table \arabic{num}: }{\usecounter{num}\setlength{\rightmargin}{\leftmargin}}

\item Estimates of two-photon and two-gluon decays widths, based upon analyses of references \cite{cf}, \cite{cfl}, and \cite{kada}. 

\item Photon-photon cross sections for $f_{J=0}(1710)$ glueball production in peripheral collisions of $^{208}$Pb at LHC and RHIC and of protons at RHIC.  For comparison, inelastic scattering effects have been accounted for in two ways: $\sigma_{AA}^{\rm el}$ uses the Glauber approximation (see equation (34)), whereas $\sigma_{AA}(b > 2R)$ is the remaining cross section after applying a cut on impact parameter.

\item Pomeron-pomeron cross sections for $f_{J=0}(1710)$ glueball production in peripheral collisions of $^{208}$Pb at LHC and RHIC and of protons at RHIC using the structure function $G_{_{\cal P}}$ of Equation (21).  For comparison, inelastic scattering effects have been accounted for in two ways: $\sigma_{AA}^{\rm el}$ uses the Glauber approximation (see equation (34)), whereas $\sigma_{AA}(b > 2R)$ is the remaining cross section after applying a cut on impact parameter.
\end{list}

\pagebreak
\begin{center}
\begin{tabular}{| c || c | c | }
\hline
 State & ~$\Gamma_{\gamma\gamma}$~ & ~$\Gamma_{gg}$~ \\ \hline\hline
$~(q\bar{q})~$ &~$\sim 1$ keV~ & $\sim$ 28 MeV~\\ 
$(G)$ & $\sim 4$ eV & \,\raisebox{-0.13cm}{$\stackrel{\textstyle>}{\textstyle\sim}$}\, 70 MeV~ \\ 
 \hline
\end{tabular}

\vspace*{.25in}
Table 1
\end{center}

\vspace*{1.0in}

\begin{center}
\begin{tabular}{|c||c|c|c|}
\multicolumn{1}{c}{}
 &\multicolumn{2}{c}{$\hbox{Pb-Pb}$}   
 &\multicolumn{1}{c}{$pp$} \\ \hline
 ~~~~$\sigma_{AA}$~~~~& ~{$\hbox{3.5 TeV/u} \atop \hbox{\small (LHC)}$}  ~  
                       & ~{$\hbox{100 GeV/u} \atop \hbox{\small (RHIC)}$}    ~
                       &~ {$\hbox{250 GeV} \atop \hbox{\small (RHIC)}$}~   \\ \hline\hline
Total&~~2.5 $\mu$b~~&~~85 nb~~&~~0.51 pb~~\\ 
Elastic &2.1 $\mu$b & 37 nb & 0.48 pb  \\ 
$b>2R$ &2.3 $\mu$b & 48 nb & 0.43 pb \\ \hline
\end{tabular}

\vspace*{.25in}
Table 2
\end{center}

\vspace*{1.0in}

\begin{center}
\begin{tabular}{|c||c|c|c|}
\multicolumn{1}{c}{}
 &\multicolumn{2}{c}{$\hbox{Pb-Pb}$}   
 &\multicolumn{1}{c}{$pp$} \\ \hline
 ~~~~$\sigma_{AA}$~~~~&~ {$\hbox{3.5 TeV/u} \atop \hbox{\small (LHC)}$} ~   
                       &~ {$\hbox{100 GeV/u} \atop \hbox{\small (RHIC)}$} ~   
                       & ~{$\hbox{250 GeV} \atop \hbox{\small (RHIC)}$}  ~ \\ \hline\hline
Total&~~2160 b~~&73 b&~~151 $\mu$b~~\\ 
Elastic & 95 mb &~~8.2 mb~~& 121 $\mu$b \\ 
$b>2R$ &15 b & 460 mb & 27 $\mu$b \\ \hline
\end{tabular}

\vspace*{.25in}
Table 3
\end{center}

\pagebreak

\section*{Figures}
%\large

\newcounter{numb}
\begin{list}%
{FIG. \arabic{numb}: }{\usecounter{numb}\setlength{\rightmargin}{\leftmargin}}

\item The fusion of two quanta $V$, $V=\gamma$ or $\cal P$, from scattering nuclei.

\item Inclusive glueball production, ${\cal PP} \rightarrow G + X$.

\item The impact parameter dependence of the differential cross section for (a) ${\cal PP}$ and (b) $\gamma\gamma\rightarrow f_{J=0}(1710)$ in peripheral collisions of $^{208}$Pb at RHIC.  The solid curves are without absorption whereas the dotted  curves include absorption effects.  

\item The impact parameter dependence of the differential cross section for ${\cal PP}\rightarrow f_{J=0}(1710)$ in proton-proton scattering at RHIC.  The solid curve is without absorption whereas the dotted curve includes absorption effects.  

\item Comparison of  different structure functions on total production cross sections for  ${\cal PP}\rightarrow f_{J=0}(1710)$ in peripheral collisions of $^{208}$Pb at (a) the LHC and (b) RHIC, and (c) for $pp$ collisions at RHIC.  The three structure functions are $G_{_{\cal P}}$ (solid), $G^{\prime}_{_{\cal P}}$ (dashed) , and $G^{\prime\prime}_{_{\cal P}}$ (dotted) of Equations (21)-(23).   

\item The ratio of ${\cal PP}$ to $\gamma\gamma$ cross sections for $f_{J=0}(1710)$ production in peripheral collisions of $^{208}$Pb at RHIC. 

\end{list}

\end{document}